\documentclass[prd,aps,showpacs,tightenlines,nofootinbib,psfig,preprint]{revtex4}
\usepackage{graphicx}

\newcommand{\CO}{{\cal O}} 
 \newcommand{\CR}{{\cal R}}

\newcommand{\bear}{\begin{array}}  \newcommand{\eear}{\end{array}}
\newcommand{\bea}{\begin{eqnarray}}  \newcommand{\eea}{\end{eqnarray}}
\newcommand{\beq}{\begin{equation}}  \newcommand{\eeq}{\end{equation}}
\newcommand{\bef}{\begin{figure}}  \newcommand{\eef}{\end{figure}}
\newcommand{\bec}{\begin{center}}  \newcommand{\eec}{\end{center}}
  
\newcommand{\lmk}{\left(}  \newcommand{\rmk}{\right)}
\newcommand{\lkk}{\left[}  \newcommand{\rkk}{\right]}
  
\newcommand{\lnk}{\left \{ }  \newcommand{\rnk}{\right \} }

\newcommand{\bib}{\bibitem}

\newcommand{\ns}{n_s}
\newcommand{\Mpc}{{\rm Mpc}}
\newcommand{\mg}{M_G}

\newcommand{\phidot}{\dot{\phi}}
\newcommand{\Psibar}{\overline{\Psi}}
\newcommand{\muni}{\mu^2}

\newcommand{\kn}{c_N}
\newcommand{\order}{\CO}
\newcommand{\Phimin}{\Phi_{\min}}
\newcommand{\calr}{{\cal R}}

\def\IBID#1#2#3{{\it ibid}. {\bf #1}, #2 (19#3)}

\def\JL#1#2#3{JETP. Lett. {\bf #1}, #2 (19#3)}

\def\PLB#1#2#3{Phys. Lett. B {\bf #1}, #2 (19#3)}

\def\PRD#1#2#3{Phys. Rev. D {\bf #1}, #2 (19#3)}

\begin{document}
\title{Inflation with a running spectral index in supergravity}
\author{M. Kawasaki}
\affiliation{Research Center for the Early Universe, University of
 Tokyo, Tokyo, 113-0033, Japan}

\author{Masahide Yamaguchi} 
\affiliation{Physics Department, Brown University, Providence, Rhode Island
  02912}
\author{Jun'ichi Yokoyama} 
\affiliation{Department of Earth and Space Science, Graduate School of
  Science, Osaka University, Toyonaka 560-0043, Japan}

\date{\today}


\begin{abstract}
  The first year Wilkinson Microwave Anisotropy Probe data favors
  primordial adiabatic fluctuation with a running spectral index with
  $\ns >1$ on a large scale and $\ns <1$ on a smaller scale.  The
  model building of inflation that predicts perturbations with such a
  spectrum is a challenge, because most models predict fluctuations
  with either $\ns >1$ (hybrid inflation) or $\ns <1$ (new, chaotic,
  or topological inflation). We give a sensible particle physics model
  in supergravity that accommodates the desired running of the
  spectral index using double inflation.
\end{abstract}

\pacs{98.80.Cq,04.65.+e,11.27.+d \hspace{1.5cm}
RESCEU-10/03, BROWN-HET-1352, OU-TAP-202}

\maketitle


\section{Introduction}

The first year data release of the Wilkinson Microwave Anisotropy Probe
(WMAP) has opened a new era of high-precision cosmology
\cite{Bennett:2003bz}.  It has not only confirmed the ``concordance''
values of the cosmological parameters with much smaller uncertainties
than before but also extracted important information on the
primordial spectrum of density perturbations \cite{Spergel:2003cb}.
It has been reported that \cite{Peiris:2003ff} their result favors
 purely adiabatic fluctuations with a remarkable
feature that the spectral index runs from $\ns>1$ on a large scale to
$\ns<1$ on a smaller scale.  More specifically they obtain
$\ns=1.10^{+0.07}_{-0.06}$ and $d\ns/d\ln k=-0.042^{+0.021}_{-0.020}$
on the scale $k_0=0.002\,\Mpc^{-1}$.

It is not straightforward to make a model of inflation \cite{oriinf}
that predicts perturbations with such a spectrum. In a single-field
slow-roll inflation model with a potential $V[\phi]$, the amplitude of
curvature perturbation in the comoving gauge $\calr$ \cite{Bardeen}
generated on the comoving scale $r=2\pi/k$ is given by
\beq
 \calr (k)=\frac{1}{2\pi}\frac{H^2(t_k)}{|\phidot(t_k)|},~~~
 H^2(t_k)=\frac{V[\phi(t_k)]}{3\mg^2},
 \label{eq:fluc}
\eeq
where $t_k$ is the epoch $k$ mode left the Hubble radius during
inflation \cite{pert}.  The spectral index is defined and given by
\beq
  \ns-1 \equiv \frac{d\ln |\calr (k)|^2}{d\ln k}
  \cong -6\epsilon+2\eta,~~~ \epsilon\equiv \frac{\mg^2}{2}
         \lmk\frac{V'[\phi]}{V[\phi]}\rmk^2,~~~\eta\equiv
         \frac{\mg^2V''[\phi]}{V[\phi]}.
  \label{eq:index}
\eeq
We also find
\beq
  \frac{d\ns}{d\ln k}=16\epsilon\eta-24\epsilon^2-2\xi,~~~\xi\equiv
 \mg^4\frac{V'''[\phi]V'[\phi]}{V[\phi]^2}.
\eeq
From Eqs. (\ref{eq:index}) we find that $\eta$ or the effective mass
of the inflaton $\phi$ must change significantly to achieve running of
the spectral index from $\ns>1$ to $\ns<1$.\footnote{Recently, Feng
{\it et al.} discussed an inflation model in the context of an extra
dimension and showed that a large variation of spectral index can be
realized \cite{FLZZ}.} Here $\mg=2.4\times 10^{18}$GeV is the reduced
Planck mass.  Hereafter we take $\mg=1$.

One possibility to realize such a spectral feature is a hybrid
inflation model of Linde and Riotto, in which supergravity effect
becomes dominant in the early stage of inflation and one-loop effect
plays an important role in the final stage of inflation
\cite{Linde:1997sj}. Unfortunately, however, although the spectral
index crosses unity in this model, its variation is too mild to
reproduce the WMAP result for a sensible set of model
parameters.\footnote{Recently, Kyae and Shafi discussed hybrid
inflation in four- and five- dimensional grand unified theories. But
it is shown that the running of the spectral index is too mild also in
this model \cite{KS}.} This is mainly because the Yukawa coupling
constant must be relatively small for sufficient inflation as we argue
below.

In order to circumvent this problem we consider a double inflation
model combining hybrid inflation \cite{Linde:1993cn} and new inflation
\cite{ni} in supergravity. In our model the first stage of hybrid
inflation proceeds in the same way as the Linde-Riotto model
\cite{Linde:1997sj}, which is a supergravity version of a
supersymmetric model \cite{Copeland:1994vg, Dvali:ms}, and the desired
spectrum with a running index from $\ns >1$ to $\ns < 1$ is generated
then.  At the same time the initial condition for subsequent new
inflation is naturally prepared \cite{Izawa:1997df}, which continues
about 40 $e$-folds to expand the comoving scale with the desired
spectral shape of density fluctuation to larger scales observable with
WMAP.  Furthermore, new inflation predicts sufficiently a low
reheating temperature to avoid overproduction of gravitinos.

The rest of the paper is organized as follows. In Sec. II we review
the supergravity hybrid inflation model of Linde and Riotto
\cite{Linde:1997sj} and argue its difficulty to realize a
significantly large running of the spectral index on scales probed by
WMAP.  Then in Sec. III we introduce our double inflation model which
was originally proposed in \cite{Izawa:1997df} and utilized to realize
the formation of primordial black holes \cite{Kawasaki:1998vx} or to
fit observations of large-scale structure
\cite{Kanazawa:1999ag}. Section IV is devoted to a discussion and
future outlook.

\section{Hybrid inflation in supergravity}

Here we first review the supergravity hybrid inflation model of Linde
and Riotto \cite{Linde:1997sj}.  The superpotential is given by
\beq
  W_H=\lambda S\Psibar\Psi -\muni S , \label{Hsuper}
\eeq
where $S$ is a gauge-singlet superfield, while $\Psi$ and $\Psibar$
are a conjugate pair of superfields transforming as nontrivial
representations of some gauge group, and $\lambda$ and $\mu$ are
positive parameters much smaller than unity.  This system respects the
$R$ symmetry under which they are transformed as $S\longrightarrow
e^{2i\alpha}S$, $\Psi \longrightarrow e^{2i\alpha}\Psi$, $\Psibar
\longrightarrow e^{-2i\alpha}\Psibar$, and $W_H \longrightarrow
e^{2i\alpha}W_H$. The $R$-invariant K\"ahler potential is given by
\beq
 K_H=|S|^2+|\Psi|^2+|\Psibar|^2+\cdots . 
\label{Hkaehler}
\eeq
Here, for simplicity, we have neglected higher-order terms, which make
the calculation complicated but do not change the essential result.

The potential of scalar components of the superfields $z_{i}$ in
supergravity is given by
\beq
  V = e^{K} \left\{ \left(
      \frac{\partial^2K}{\partial z_{i}\partial z_{j}^{*}}
    \right)^{-1}D_{z_{i}}W D_{z_{j}^{*}}W^{*}
    - 3 |W|^{2}\right\} + V_{D},
  \label{Gpotential}
\eeq
with 
\beq
  D_{z_i}W = \frac{\partial W}{\partial z_{i}} 
    + \frac{\partial K}{\partial z_{i}}W.
  \label{DW}
\eeq
Here the scalar components of the superfields are denoted by the same
symbols as the corresponding superfields, and $V_{D}$ represents the
$D$-term contribution. From Eqs. (\ref{Hsuper}) and (\ref{Hkaehler})
we explicitly find
\bea
 V_H(S,\Psi,\Psibar)&=&e^{|S|^2+|\Psi|^2+|\Psibar|^2}
\lnk (1-|S|^2+|S|^4)
|\lambda\Psibar\Psi-\muni |^2 \right.\nonumber \\
& &\left.+|S|^2\lkk\left|\lambda(1+|\Psi|^2)\Psibar-\muni\Psi^\ast \right|^2
 +\left|\lambda(1+|\Psibar|^2)\Psi-\muni\Psibar^\ast \right|^2\rkk\rnk
+V_{DH}. \label{Hpotential}
\eea
Since the above potential does not depend on the phase of the complex
scalar field $S$, we identify its real part, $\sigma\equiv {\rm
Re}S/\sqrt{2}$, with the inflaton. Composing the mass matrix of $\Psi$
and $\Psibar$ from Eq. (\ref{Hpotential}), we find that the
eigenvalues and the corresponding eigenstates are given by
\beq
M^2_\pm=\lambda^2|S|^2\pm\lambda\muni
=\frac{\lambda^2}{2}\sigma^2\pm\lambda\muni,
~~~{\rm for}~~~\Psi^\ast=\mp \Psibar,
\eeq
where we have assumed $\lambda \gg \mu$.  Since these eigenstates are
along $D$-flat directions, the scalar fields will eventually start to
roll, keeping $\Psi^\ast= \Psibar$, as $\sigma$ becomes smaller than the
critical value $\sigma_c\equiv \sqrt{2}\mu/\sqrt{\lambda} $.

Then under the condition $\Psi^\ast=\Psibar$, the potential reduces to
a familiar form of the typical hybrid inflation potential:
\beq
 V_H=(\lambda|\Psi|^2-\muni)^2+\lambda^2\sigma^2|\Psi|^2+
 \frac{1}{8}\mu^4\sigma^4+\cdots.
\eeq
For $\sigma > \sigma_c$, the potential is minimized at
$\Psi=\Psibar=0$ and inflation is driven by the false vacuum energy
density $\mu^4$.

As a result of the mass split of scalar multiplets composed by $\Psi$
and $\Psibar$ with mass squared $M^2_\pm$ and their superpartner
fermions with mass $M=\lambda\sigma/\sqrt{2}$, the radiative
correction to the potential is non-negligible during inflation.  The
one-loop correction reads~\cite{Dvali:ms}
\beq
 V_{1L}=\frac{\lambda^2}{128\pi^2}\lkk (\lambda\sigma^2+2\muni)^2
 \ln\frac{\lambda\sigma^2+2\muni}{\Lambda^2}
+(\lambda\sigma^2-2\muni)^2
 \ln\frac{\lambda\sigma^2-2\muni}{\Lambda^2}
-2\lambda^2\sigma^4\ln\frac{\lambda\sigma^2}{\Lambda^2}\rkk,
\eeq
where $\Lambda$ is the renormalization scale.  When $\sigma \gg
\sigma_c$ it is approximated as
\beq
 V_{1L}\cong \frac{\lambda^2\mu^4}{8\pi^2}\ln\frac{\sigma}{\sigma_c}.
 \label{Lapprox}
\eeq
As a result the effective potential of the inflaton $\sigma$ during
hybrid inflation reads
\beq
 V_H[\sigma]\cong \mu^4\lmk 1+\frac{\lambda^2}{8\pi^2}
\ln\frac{\sigma}{\sigma_c} +\frac{1}{8}\sigma^4 +\cdots\rmk. 
\label{Hpotea}
\eeq

As long as the field amplitude is sub-Planck scale or smaller than
unity, the effective potential is dominated by the false vacuum
energy.  Comparing the derivative of the second term of the right-hand
side and that of the last term, we find that the dynamics of the
scalar field is dominated by the nonrenormalizable term for $\sigma >
\sqrt{\lambda/(2\pi)}\equiv\sigma_d$ and by the radiative correction
for $\sigma < \sigma_d$.  Since we are assuming $\mu \ll \lambda \ll
1$, we find $\sigma_c \ll \sigma_d \ll 1$.  In this situation, the
number of $e$-folds of exponential expansion during inflation, $N_H$,
is sensitive only to $\sigma_d$ and given by
\beq
 N_H=\int_{\sigma_c}^{\sigma_i}\frac{V_H[\sigma]}{V_H'[\sigma]}d\sigma
 \cong \int_{\sigma_d}^{\sigma_i}\frac{2}{\sigma^3}d\sigma
 +\int_{\sigma_c}^{\sigma_d}\frac{8\pi^2}{\lambda^2}\sigma d\sigma
 \cong \frac{4\pi}{\lambda}, \label{Hefold}
\eeq 
where $\sigma_i$ is the initial value at the onset of hybrid inflation.
Although the approximate expression for the one-loop potential
(\ref{Lapprox}) is valid only for $\sigma \gg \sigma_c $, the fact
that the resultant expression of $N_H$ is independent of $\sigma_c$
implies that the above formula (\ref{Hefold}) gives a reasonable
approximation for the actual number of $e$-folds.  We also note that
the amount of inflation during $\sigma>\sigma_d$ and that during
$\sigma<\sigma_d$ are about the same with the $e$-folds $\approx
2\pi/\lambda$.  Thus in order to achieve sufficiently long inflation
to solve the horizon and the flatness problems, $N_H \gtrsim 60$,
$\lambda$ should be rather small: $\lambda \lesssim 4\pi/60 \simeq
0.2$.

Calculating the slow-roll parameters $\epsilon$, $\eta$, and $\xi$ for
the potential (\ref{Hpotea}), we find 
\beq
 \epsilon=\order\lmk\lmk\frac{\sigma}{\mg}\rmk^6\rmk,~~
 \eta=\order\lmk\lmk\frac{\sigma}{\mg}\rmk^2\rmk,~~
 \xi=\order\lmk\lmk\frac{\sigma}{\mg}\rmk^4\rmk,
\eeq
for $\sigma\sim\sigma_d$ in dimensionful units, so that the spectral
index of scalar perturbation and its variation are given by
\bea
 \ns-1&=&-6\epsilon+2\eta\cong
 2\eta=3\sigma^2-\frac{\sigma_d^4}{\sigma^2}, \label{ns} \\
 \frac{d\ns}{d\ln k}&=& 16\epsilon\eta-24\epsilon^2-2\xi
 \cong -2\xi=-\lmk\frac{\sigma_d^8}{\sigma^4}+4\sigma_d^4+3\sigma^4\rmk.
 \label{run}
\eea
As is seen here, this model has the desired feature of the running
spectral index toward smaller values for decreasing length scales
qualitatively.  The spectral index crosses unity at
$\sigma=\sigma_d/3^{1/4}\sim 0.8\sigma_d$.  In order that this happens
on the observable scales by WMAP and large-scale structures, one
should have about 50 $e$-folds of inflation after $\sigma$ crossed
this value, which requires $\lambda\sim 0.1$ from Eq. (\ref{Hefold}).
As a result, we find that the amplitude of spectral running is too
small:
\beq
  \frac{d\ns}{d\ln k}\sim -10^{-3}.
\eeq

Conversely, this quantity could be larger and match observation if we
adopted a larger value of the coupling constant $\lambda$.  Indeed
from Eqs. (\ref{ns}) and (\ref{run}) we find \bea \sigma_d
&=&\frac{1}{2\sqrt{2}}\lnk\lkk \frac{4}{9}(\ns-1)^4
-\frac{32}{3}(\ns-1)^2\frac{d\ns}{d\ln k}\rkk^{\frac{1}{2}}
-\frac{14}{3}(\ns-1)^2-8\frac{d\ns}{d\ln k}\rnk^{\frac{1}{4}}, \\
\sigma &=& \lkk \frac{\ns-1}{2}+\frac{1}{2(\ns-1)}\frac{d\ns}{d\ln k}
+\frac{4\sigma_d^4}{\ns-1}\rkk^{\frac{1}{2}}, \eea where $\sigma$ is
the field amplitude when the relevant scale left the Hubble radius
during hybrid inflation.  Using the central values obtained by WMAP on
a comoving scale, $k_0=0.002\,{\rm Mpc}^{-1}$, $\ns-1=0.1$ and
$d\ns/d\ln k=-0.042$, we find $\sigma_d=0.27$ or $\lambda=0.47$ and
$\sigma=0.25=0.91\sigma_d$.  This means that the $e$-folds of hybrid
inflation after the comoving scale with the observed spectral shape
has crossed the Hubble radius is only about 11.3. Hence we must invoke
another inflation to push the relevant scale to the scale
$k_0=0.002\,{\rm Mpc}^{-1}$.  If this is achieved, we can find the
energy scale of hybrid inflation from the amplitude of curvature
fluctuation at $\sigma=0.25$,
\beq
 \calr(k_0)=\frac{\muni}{\sqrt{3}\pi}
\lmk\frac{\sigma_d^4}{\sigma}+\sigma^3\rmk^{-1} 
=4.9\muni=4.8\times10^{-5},
\eeq
corresponding to $A=0.77$ of \cite{Peiris:2003ff}.  We therefore find
$\mu=3.1\times 10^{-3}=7.5\times 10^{15}$GeV.

Although we have taken the central values obtained by WMAP, the
$n_s-1$ and $d\ns/d\ln k$ are determined with rather large
uncertainties. For example, if we take $d\ns/d\ln k = -0.021$
corresponding to 68\% probability lower bound, we obtain $\sigma_d =
0.22 (\lambda=0.30, \sigma = 0.95\sigma_d)$, which changes the
$e$-folds of the hybrid inflation from $11.3$ to $19.9$. Thus, we
should take $\Delta N_H \simeq 8$ as the uncertainty of $e$-folds of
hybrid inflation.

\begin{figure}
\includegraphics[width=16cm]{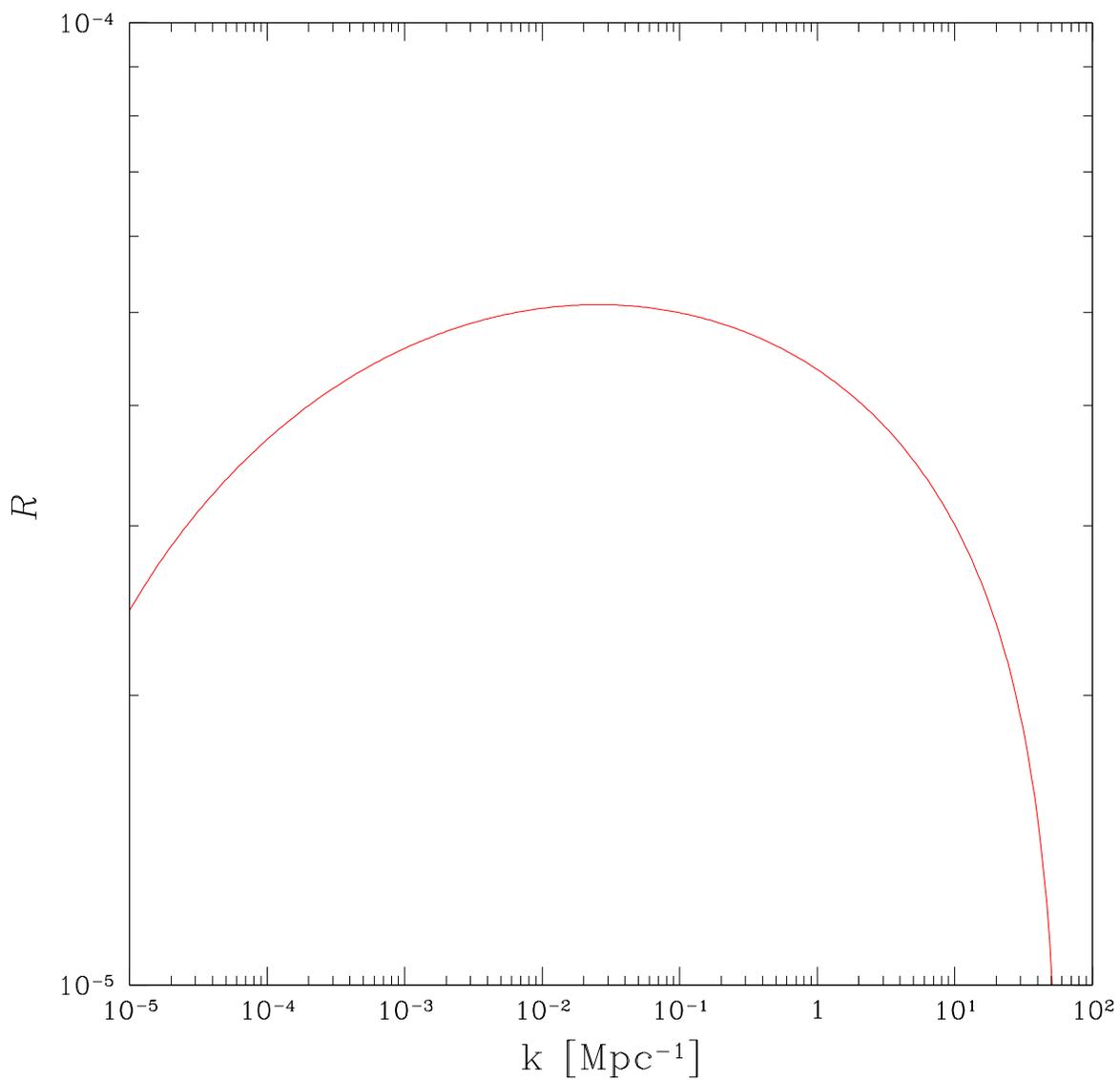} 
\caption{\label{fig:spectrum} 
  The amplitude of the curvature fluctuation $\CR$ during hybrid
  inflation is shown with the parameters $\mu = 2.9 \times 10^{-3},
  \lambda = 0.43$. The result is independent of $v$ and $g$ as long as
  $v \ll \mu$.}
\end{figure}

\begin{figure}
\includegraphics[width=16cm]{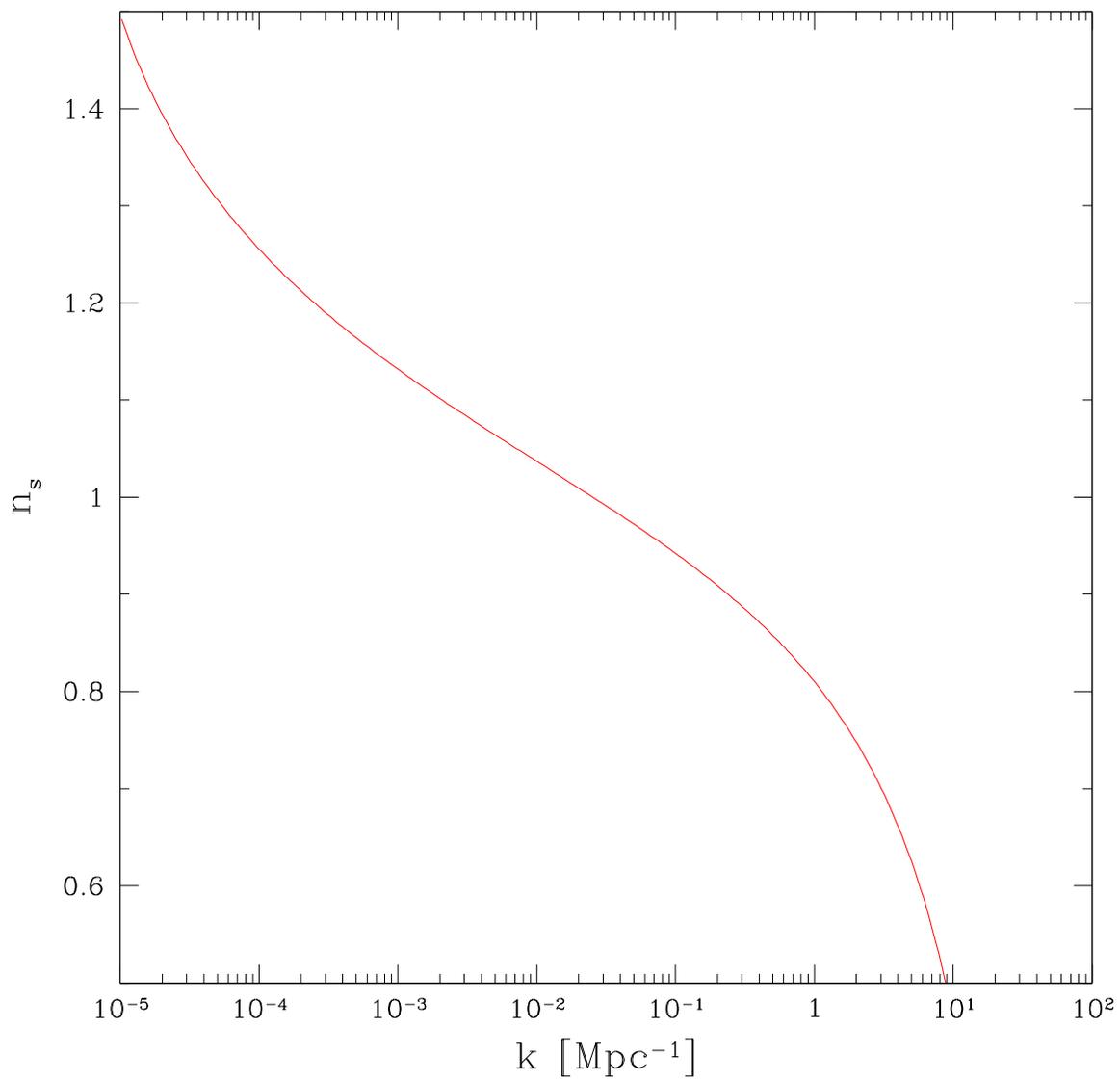} 
\caption{\label{fig:index} 
  The spectral index $n_s$ during hybrid inflation is shown with the
  same parameters.}
\end{figure}

\begin{figure}
\includegraphics[width=16cm]{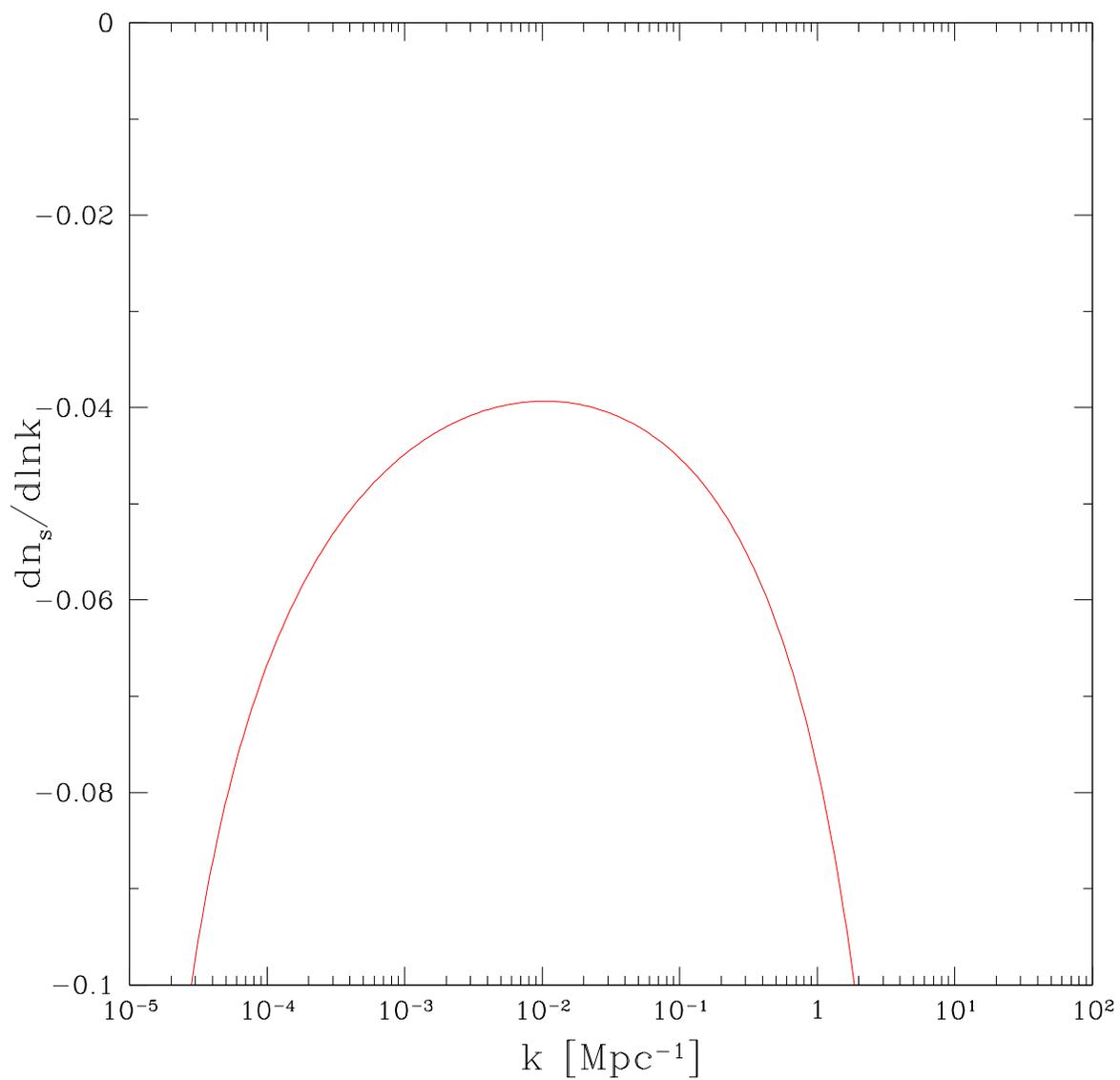} 
\caption{\label{fig:derindex} 
  The derivative of the spectral index $d\ns/d\ln k$ during hybrid
  inflation is shown with the same parameters.}
\end{figure}

Note also that the two contributions from the supergravity effect and
the one-loop correction are dealt with separately in the above
analytic estimate. However, of course, both contributions should be
considered simultaneously.  So we have calculated the fluctuations
numerically. Then, we have found that the best-fit parameters to
reproduce the WMAP results are slightly changed and given by $\mu =
2.9 \times 10^{-3} = 7.2 \times 10^{15}$ GeV, $\lambda = 0.43$. The
results are shown in Figs.\ \ref{fig:spectrum}-\ref{fig:derindex}. In
this case, the $e$-folds of hybrid inflation after the comoving scale
with the observed spectral shape has crossed the Hubble radius is only
about 10.3.

\section{Hybrid new inflation}

In order to push the scales with the desired spectral shape of density
fluctuations to cosmologically observable scales, we consider new
inflation which occurs after hybrid inflation discussed above. One
should notice that, generally speaking, hybrid inflation predicts a
high reheating temperature because the inflaton has gauge couplings
and its energy scale is relatively high usually. On the other hand,
new inflation predicts a sufficiently low reheating temperature to
avoid overproduction of gravitinos. Thus, the occurrence of new
inflation following hybrid inflation is favorable also in this
respect. Furthermore, in our model, the initial value of new inflation
is dynamically set during hybrid inflation, which evades the severe
initial value problem of new inflation. In this section, we first
review the superpotential and K\"ahler potential for the new inflation
sector and its dynamics \cite{Izawa:1997df}, and then combine these
two inflation by considering full superpotential.

We introduce a chiral superfield $\Phi$ with an $R$ charge $2/(n+1)$,
but assume that the U$(1)_R$ symmetry is dynamically broken to a
discrete $Z_{2n~R}$ symmetry at a scale $v \ll 1$.  The superpotential
of this sector therefore reads
\beq
 W_N[\Phi] = v^2\Phi - \frac{g}{n+1}\Phi^{n+1}, \label{Nsuper}
\eeq
with $g$ being a coupling constant of order of unity.  We assume
that both $g$ and $v$ are real and positive for simplicity.  The
$R$-invariant K\"ahler potential is given by
\beq
 K_N=|\Phi|^2+\frac{\kn}{4}|\Phi|^4+\cdots, \label{Nkaehler}
\eeq
where $\kn$ is a constant smaller than unity.  We assume that there
are no direct interactions between fields relevant to hybrid inflation
discussed in the previous section and $\Phi$.  Hence the full
superpotential and K\"ahler potential read $W=W_H+W_N$ and
$K=K_H+K_N$, respectively.

Before performing the full analysis we review how new inflation
proceeds in this model.  The scalar potential given from
Eqs. (\ref{Nsuper}) and (\ref{Nkaehler}) reads
\bea
V_N[\Phi]&=&\frac{\exp\lmk |\Phi|^2+\frac{\kn}{4}|\Phi|^4\rmk}
{1+\kn|\Phi|^2} \nonumber \\
& &\times\lkk~ \left| \lmk 1+|\Phi|^2+\frac{\kn}{2}|\Phi|^4\rmk v^4
-\lmk 1+\frac{|\Phi|^2}{n+1}+\frac{\kn|\Phi|^4}{2(n+1)}\rmk
g\Phi^n \right|^2 \right. \nonumber \\ 
& &~~~\left.
-3\lmk 1+\kn|\Phi|^2 \rmk|\Phi|^2\left| v^2-\frac{g}{n+1}\Phi^n
\right|^2\rkk .
\eea
It has a minimum at
\beq
 |\Phi|_{\min}\cong\lmk\frac{v^2}{g}\rmk^{\frac{1}{n}}
 ~~~{\rm and~~Im}\Phi^n_{\min}=0, 
\eeq
with a negative energy density
\beq
 V_N[\Phimin]\cong -3e^{K_N}|W_N[\Phimin]|^2\cong
 -3\lmk\frac{n}{n+1}\rmk^2v^4|\Phimin|^2.
\eeq
Assuming that this negative value is canceled by a positive
contribution due to supersymmetry breaking, $\Lambda_{\rm SUSY}^4$, we
can relate energy scale of this model with the gravitino mass
$m_{3/2}$ as
\beq
 m_{3/2}\cong \frac{n}{n+1}\lmk\frac{v^2}{g}\rmk^{\frac{1}{n}}v^2.
\eeq

Without loss of generality we may identify the real part of $\Phi$
with the inflaton $\phi\equiv{\rm Re}\Phi/\sqrt{2}$.  The dynamics of
inflaton is governed by the lower-order potential
\beq
 V_N[\phi]\cong v^4-\frac{\kn}{2}v^4\phi^2-\frac{2g}{2^{n/2}}v^2\phi^n 
 +\frac{g^2}{2^n}\phi^{2n}. \label{Neffpote}
\eeq
Since the last term is negligible during inflation and the Hubble
parameter is dominated by the first term, $H=v^2/\sqrt{3}$, the
slow-roll equation of motion reads
\beq
 3H\dot{\phi}=-V'_N[\phi]\cong
-\kn v^4\phi-{2^{\frac{2-n}{2}}}{ngv^2}
\phi^{n-1},  \label{Neqm}
\eeq
and the slow-roll parameters are given by
\beq
 \epsilon\cong\frac{1}{2}\lmk\kn\phi+{2^{\frac{2-n}{2}}}{ng}
\frac{\phi^{n-1}}{v^2}\rmk^2,~~~
\eta=-\kn-{2^{\frac{2-n}{2}}}{n(n-1)g}\frac{\phi^{n-2}}{v^2},
\eeq
in this new inflation regime.  Thus inflation is realized with $\kn
\ll 1$ and ends at
\beq
  \phi=\sqrt{2}\lmk\frac{(1-\kn)v^2}{gn(n-1)}\rmk^{\frac{1}{n-2}}\equiv
  \phi_e,
\eeq
when $|\eta|$ becomes as large as unity.

Since the two terms on the right-hand side of Eq. (\ref{Neqm}) are
identical at
\beq
  \phi=\sqrt{2}\lmk\frac{\kn v^2}{gn}\rmk^{\frac{1}{n-2}}\equiv\phi_d,
\eeq
the number of $e$-folds of new inflation is estimated as
\beq
 N_N=-\int_{\phi_i}^{\phi_e}\frac{V_N[\phi]}{V'_N[\phi]}d\phi
 \cong\int_{\phi_i}^{\phi_d}\frac{d\phi}{\kn\phi} +
\int_{\phi_d}^{\phi_e}\frac{2^{\frac{n-2}{2}}v^2}{gn\phi^{n-1}}d\phi
=\frac{1}{\kn}\ln\frac{\phi_d}{\phi_i}+\frac{1-n\kn}{(n-2)\kn(1-\kn)},
\eeq
for $0<\kn<n^{-1}$.
If $\kn$ vanishes, we instead find
\beq
 N_N=\int_{\phi_i}^{\phi_e}
\frac{2^{\frac{n-2}{2}}v^2}{gn\phi^{n-1}}d\phi
=\frac{2^{\frac{n-2}{2}}v^2}{gn(n-2)}
\phi_i^{2-n}-\frac{n-1}{n-2}. \label{nnzero}
\eeq
Here $\phi_i$ is the initial value of $\phi$, whose determination
mechanism we now argue.

In the hybrid inflation stage $\sigma > \sigma_c$, the cosmic energy
density is dominated by the false vacuum energy $\mu^4$ with
$\Psi=\Psibar=0$. Hence, in the full scalar potential, which is
obtained by the prescription (\ref{Gpotential}) with $K=K_H+K_N$ and
$W=W_H+W_N$, the interaction terms between $\Phi$ and the hybrid
inflation sector are given by
\beq
 V \supset \mu^4|\Phi|^2+\muni v^2 (\Phi^\ast S+\Phi S^\ast)+\cdots
=\frac{1}{2}\mu^4(\phi^2+\chi^2)+\muni
v^2\sigma\phi+\cdots,
\eeq
where $\phi={\rm Re}\Phi/\sqrt{2}$ and $\chi={\rm Im}\Phi/\sqrt{2}$.
Hence at the end of hybrid inflation, $\sigma=\sigma_c$, $\phi$ and
$\chi$ have a minimum at
\beq
 \phi_{\min}\cong-\frac{v^2}{\muni}\sigma_c
= -\sqrt{\frac{2}{\lambda}}\frac{v^2}{\mu}
~~~{\rm and}~~~ \chi_{\min}\cong 0,
\label{eq:minimum}
\eeq
respectively.  Since the effective mass is larger than the Hubble
parameter during hybrid inflation, the above configuration is realized
with the dispersion
\beq
 \langle(\phi-\phi_{\min})^2\rangle=\langle\chi^2\rangle
=\frac{3}{8\pi^2}\frac{H_H^4}{\mu^4}=\frac{\mu^4}{24\pi^2},
\eeq
due to quantum fluctuations \cite{BD}. Here $\phi$ will eventually
relax to this minimum during hybrid inflation.  The ratio of quantum
fluctuation to the expectation value should satisfy
\beq
 \frac{\sqrt{\langle(\phi-\phi_{\min})^2\rangle}}{|\phi_{\min}|}
=\frac{\sqrt{\lambda}}{4\sqrt{3}\pi}\lmk\frac{\mu}{v}\rmk^2\mu
\ll 1, 
\label{ratio}
\eeq
so that the initial value of the inflaton for new inflation is located
off the origin with an appropriate magnitude.

When $\sigma$ reaches $\sigma_c$, a phase transition takes place and
hybrid inflation ends immediately, which is followed by field
oscillation of $S$ and $\Psi(\Psibar)$ and their decay.  If this
oscillation phase lasts for a prolonged period due to gravitationally
suppressed interactions of these fields, $\phi$ will also oscillate
and its amplitude decreases with an extra factor $v/\mu$
\cite{Kawasaki:1998vx}.  In this case, new inflation could start with
a even smaller value of $\phi$ depending on its phase of oscillation
at the onset of inflation (see \cite{Kanazawa:1999ag} for an analytic
estimate of the initial phase).
So we set the initial value of $\phi$ as
\beq
\phi_i = \sqrt{\frac{2}{\lambda}}\frac{v^3}{\mu^2}
\eeq
and new inflation occurs until $\phi=\phi_e$ with the potential
(\ref{Neffpote}).  Thus we can understand the evolution of the
universe throughout double inflation.

Contrary to the hybrid inflation regime we do not have much precise
observational constraints on the new inflation regime, so we cannot
fully specify values of the model parameters for new inflation. Hence
let us content ourselves with a few specific examples. First we
consider the cases with $\kn=0$. Then from Eq. (\ref{nnzero}) the
number of $e$-folds of new inflation reads
\beq
 N_N=\frac{\lambda^{\frac{n-2}{2}}\mu^{2n-4}v^{8-3n}}{n(n-2)g}
-\frac{n-1}{n-2}
\cong \frac{\lambda^{\frac{n-2}{2}}\mu^{2n-4}v^{8-3n}}{n(n-2)g}. \label{nnn}
\eeq
This should be around $40 + (2/3)\ln(\mu/v)$ to push 
the comoving scale with appropriate spectral
shape to the appropriate physical length scale.\footnote{
  Comoving scales that left the Hubble radius in the late stage of
 hybrid inflation reenter the horizon before the beginning of the new
 inflation. Hence extra $e$-folds $(2/3)\ln(\mu/v)$ should be added in
 making a correspondence between comoving horizon scales during hybrid
 inflation and proper scales \cite{Kawasaki:1998vx}.  }
On the other hand, the amplitude of curvature perturbation at the
onset of new inflation, $\phi=\phi_{\min}$, is given by
\beq
  \calr = \frac{\lambda^{\frac{n-1}{2}}\mu^{2n-2}v^{7-3n}}
{2\sqrt{6}\pi n g}=\frac{n-2}{2\sqrt{6}\pi}\sqrt{\lambda}
\mu \frac{\mu}{v}N_N
\cong 4.9\times 10^{-3}(n-2)\frac{\mu}{v}\frac{N_N}{40},
\eeq
where use has been made of the values $\lambda=0.43$ and
$\mu=2.9\times 10^{-3}$ in the last equality. From Eq. (\ref{nnn}) we
find
\bea
 v&=&5.5\times 10^{-4}\lmk\frac{gN_N}{40}\rmk^{-\frac{1}{4}},~~~{\rm
for}~n=4, \nonumber \\
 v&=&4.0\times 10^{-3}\lmk\frac{gN_N}{40}\rmk^{-\frac{1}{10}},~~~{\rm
for}~n=6,  \\
 v&=&6.7\times 10^{-3}\lmk\frac{gN_N}{40}\rmk^{-\frac{1}{16}},~~~{\rm
for}~n=8.
\eea
Thus $v$ is larger than $\mu$ for $g<1$ and $n\geq 6$ in contradiction
to our assumption that new inflation takes place after hybrid
inflation at lower energy scale. For $n=4$ we find that $v$ is smaller
than $\mu$ for $g<1$ but it can be close to $\mu$ for $g\lesssim
10^{-3}$.  In this case, $\calr$ is as large as $0.01$ at the onset of
new inflation which corresponds to the comoving scale $\ell_{*}\sim
100$ kpc today.  As $g$ increases from $10^{-3}$, $\calr$ and
$\ell_{*}$ increase. For $g=1$, we find $\calr \simeq 0.05$ and
$\ell_{*} \simeq 300$ kpc.

On the other hand, for $\kn\neq 0$, we find
\beq
 \calr =\frac{\sqrt{\lambda} \mu^2}{2\pi\sqrt{6}\kn v}\cong 
  1.2\times 10^{-4}\kn^{-1}\frac{\mu}{v},
\eeq
at the onset of new inflation.  It is independent on $n$ and $g$, and
is again much larger than $10^{-5}$.  The number of $e$-folds depends
on these parameters, and for $n=4$ we find
\beq
 N_N=\frac{1}{2\kn}\ln\lmk\frac{\kn\lambda\mu^4}{4gv^4}\rmk
+\frac{1-4\kn}{2\kn(1-\kn)}.
\eeq
Thus if we take $\kn=0.1$, for example, $N_N=40$ implies $v=1.5\times
10^{-4}g^{-\frac{1}{4}}$, which corresponds to the comoving scale
$\ell_{*}\simeq 720 g^{1/6}$~kpc at the onset of new inflation.


\section{Discussion}

In the present paper we have built up a model of double inflation to
reproduce a spectrum of adiabatic fluctuations with a running spectral
index from $\ns>1$ on large scales to $\ns <1$ on smaller scales as
favored by the recent observational result of the WMAP satellite.
Such a spectrum has been known to be realized by the hybrid inflation
model in supergravity proposed by Linde and Riotto \cite{Linde:1997sj}
but we have argued that in their model running of spectral index is
too mild to be comparable with the observation. In order to cure this
problem we have shown that another inflation is necessary after hybrid
inflation and presented a specific model to realize new inflation
naturally after hybrid inflation which was originally proposed in
\cite{Izawa:1997df}. By carefully analyzing the spectrum of
fluctuations we have shown that under some natural choice of model
parameters the desired spectrum can be realized. In particular, model
parameters of hybrid inflation are fully specified from the spectral
shape and we find that the energy scale of first inflation is fairly
large, $\mu=2.9\times 10^{-3}$, which means that tensor fluctuations
can be detected by PLANCK.
On the other hand, it has been pointed out that hybrid inflation has a
problem of initial conditions \cite{ml}: namely, only a very limited
configurations of multi-field space can lead to hybrid inflation
unlike chaotic inflation.  The resolution of this problem in the
present context will be published elsewhere \cite{yy}.

A generic prediction of our double inflation model is that the
amplitude of the adiabatic fluctuations is quite large below the scale
corresponding to the horizon at the onset of new inflation ($<
\ell_{*}$).  This results in the early formation of dark halo objects
with comoving scale $\ell_{*}$.  If $\ell_{*}$ is larger than about
1~kpc, the dark halos may cause a cosmological problem because they
significantly harm subsequent galaxy formation or produce too many
gravitational lens events. Our model predicts the critical scale
$\ell_{*} \gtrsim 100$~kpc which leads to the above cosmological
problem. However, the critical scale can be taken much smaller when
the uncertainty of the WMAP data is taken into account.  As discussed
in Sec. II, the WMAP data imply that the uncertainty of the $e$-folds
for hybrid inflation is $\Delta N_H \sim 8$. Taking a little larger
$e$-folds $N_H \simeq 15$ gives $\ell_{*} < 1$~kpc, and hence the
problem can be avoided.  For this case, the predicted small dark halos
may play an important role in galaxy formation and early star
formation as inferred by WMAP~\cite{Bennett:2003bz}. This subject is
beyond the scope of the present paper and will be investigated
elsewhere.  Another prediction of our model is that contrary to the
case of single hybrid inflation, the reheating temperature is low
enough to avoid the gravitino problem.

\acknowledgements{ We are grateful to Joe Silk for useful comments.
  This work was partially supported by the JSPS Grant-in-Aid for
  Scientific Research No.\ 14540245 (M.K.), No.\ 13640285 (J.Y.), and
  Research Abroad (M.Y.). M.Y. is partially supported by the
  Department of Energy under Grant No. DEFG0291ER40688.}


\end{document}